\documentclass[aip,cha,onecolumn,superscriptaddress,preprintnumbers,showpacs, amsmath,amssymb]{revtex4-1}
\usepackage{bm}
\usepackage[colorlinks=true,linkcolor=blue,citecolor=blue]{hyperref}
\usepackage{amsmath}
\usepackage{amssymb}
\usepackage{amsthm}
\usepackage{amsfonts}
\usepackage{enumerate}
\usepackage{latexsym}
\usepackage{ifpdf}
\usepackage{graphicx}
\usepackage{makeidx}
\expandafter\ifx\csname package@font\endcsname\relax\else
 \expandafter\expandafter
 \expandafter\usepackage
 \expandafter\expandafter
 \expandafter{\csname package@font\endcsname}
 \fi
\usepackage{color}

\begin{document}

\title{Electronic and magnetic properties of (1 1 1)-oriented CoCr$_2$O$_4$ epitaxial thin film}
\author{Xiaoran Liu}
\email{xxl030@email.uark.edu}
\affiliation{Department of Physics, University of Arkansas, Fayetteville, Arkansas 72701, USA}

\author{M. Kareev} 
\affiliation{Department of Physics, University of Arkansas, Fayetteville, Arkansas 72701, USA}

\author{Yanwei Cao}
\affiliation{Department of Physics, University of Arkansas, Fayetteville, Arkansas 72701, USA}

\author{Jian Liu}
\affiliation{Department of Physics, University of California, Berkeley, California 94720, USA}
\affiliation{Materials Science Division, Lawrence Berkeley National Laboratory, Berkeley, California 94720, USA}

\author{S. Middey}
\affiliation{Department of Physics, University of Arkansas, Fayetteville, Arkansas 72701, USA}

\author{D. Meyers}
\affiliation{Department of Physics, University of Arkansas, Fayetteville, Arkansas 72701, USA}

\author{J. W. Freeland}
\affiliation{Advanced Photon Source, Argonne National Laboratory, Argonne, Illinois 60439, USA}

\author{J. Chakhalian}
\affiliation{Department of Physics, University of Arkansas, Fayetteville, Arkansas 72701, USA}

\begin{abstract}

We report on the fabrication of high quality (1 1 1)-oriented ferrimagnetic normal spinel CoCr$_2$O$_4$ epitaxial  thin films on single crystal Al$_2$O$_3$ substrates. The structural, electronic and magnetic properties were characterized by $in$-$situ$ reflection high energy electron diffraction, atomic force microscopy, X-ray diffraction, X-ray photoemission spectroscopy, SQUID magnetometry and element resolved resonant X-ray  magnetic scattering. The comprehensive characterization reveals that no disorder in the cation distribution or multivalency issue is present in the samples. As a result, Kagom$\acute{e}$ and triangular layers are naturally formed via this specific growth approach. These findings offer a pathway to fabricate  two dimensional Kagom$\acute{e}$ heterostructures with novel quantum many-body phenomena by  means of geometrical  design.  
\end{abstract}

\maketitle

\newpage

In the past few years, two dimensional (2D) Kagom$\acute{e}$ lattices have attracted tremendous interest in the pursuit of novel quantum phenomena. Many intriguing exotic quantum states on this lattice have been predicted by theoretical calculations including topological insulators\cite{Guo, Hu}, spin liquids\cite{Han, Punk}, kinetic ferromagnetism\cite{Pollmann}, spin Hall effect\cite{Liu}, anomalous Hall effect of light\cite{Petrescu}, and chiral superconducting state\cite{Yu}. However, the synthesis of a 2D Kagom$\acute{e}$ lattice is a quite challenging work. Several metal-organic hybrid compounds have been synthesized by chemical methods and possess Kagom$\acute{e}$ lattice but structural disorder and distortions remained a persistent hinderance\cite{Yunling, Emily}. There is, however,  an alternative approach to realize this situation. Recently, epitaxial thin films and heterostructures have been grown along the (1 1 1) direction and this new control parameter which is referred as geometrical engineering, has cultivated many novel systems with emergent properties\cite{Gibert, Sri, Ben}. Following  the same approach, it was pointed out that Kagom$\acute{e}$ lattice can be obtained by artificially controlling the thin film growth of a spinel type oxide (general formula AB$_2$O$_4$) along the (1 1 1) orientation\cite{Yahiro}. 

Among many choices of spinel oxides, CoCr$_2$O$_4$ (CCO) is a prototypical candidate which has been widely studied to exhibit interesting physical phenomena in the bulk materials including conical spin states\cite{Menyuk, Choi, Tomiyasu}, induced multiferroic behavior at low temperature\cite{Yamasaki, Singh}, and an unconventional magneto-structural phase transition in high magnetic fields\cite{Tsurkan}. CCO is a normal spinel oxide in which all of the Co$^{2+}$ occupies the tetrahedral sites while Cr$^{3+}$ resides within in the octahedral sites. The electronic configurations of Co$^{2+}$ and Cr$^{3+}$ are 3$\it{d}$$^7$ and 3$\it{d}$$^3$ respectively, both having S $=$ 3/2\cite{Kaplan, Menyuk, Choi, Tomiyasu}. When viewed along the (1 1 1) direction, CCO consists of alternating stacked multi-layers with Cr$^{3+}$ ions composing a Kagom$\acute{e}$ (K) lattice and a triangular (T) lattice while the Co$^{2+}$ ions composing two triangular (T') ones, forming the stacking sequence [$\cdots$T'/K/T'/T$\cdots$]. The crystal structure of the CCO conventional unit cell is depicted in Fig. 1(a). The 2D Kagom$\acute{e}$ lattice plane is shown in Fig. 1(b) with the typical Kagom$\acute{e}$ unit cell outlined by yellow lines. To date, despite the conceptual  attractiveness of such geometrical design, the  epitaxial growth of (1 1 1)  oriented spinel oxide thin films presents serious challenge due to the reported cation distribution disorder and  mixture of various transition metal oxidation states\cite{Aria, Ma, Ulrike, Matzen}. 

In this Letter, we report on the layer-by-layer growth of high quality (1 1 1)-oriented CCO thin films on Al$_2$O$_3$ (AlO) (0 0 0 1) single crystal substrates by the pulsed laser deposition technique. The crystallinity, surface morphology, film thickness and atomic structure are studied by $in$-$situ$ high pressure reflection-high-energy-electron-diffraction (RHEED), atomic force microscopy (AFM), X-ray diffraction (XRD) and X-ray reflectivity (XRR). Electronic and magnetic properties are extensively characterized by $in$-$situ$ X-ray photoemission spectroscopy (XPS), SQUID magnetometry and synchrotron-based X-ray resonant magnetic scattering (XRMS). The combined data confirm that the (1 1 1) grown CCO samples maintain a normal spinel structure with no cation distribution disorder or multivalency issue.  


CCO thin films were fabricated under a partial pressure of 5 mTorr of oxygen by the pulsed laser interval deposition method\cite{Kareev} using a KrF excimer laser operating at $\lambda$  $=$ 248 nm. During the deposition, the substrate was kept at 800 $^{\circ}$C. The laser's intensity and pulse rate were ~2 J/cm$^2$ and 18 Hz, respectively. Samples were annealed at the growth condition for 10 min and then cooled down to room temperature. Fig. 1(c) displays the RHEED pattern of the substrate before the deposition. As seen, the specular spot together with the (-1, -1) and (1, 1) spots on the Laue circle and the strongly developed Kikuchi lines testify the smooth morphology of the AlO (0 0 0 1) substrate\cite{Moussy}. During the deposition, the recovery of the intensity of the specular reflection spot and the occurrence of the half order spots indicate the 2D epitaxial growth of the film. After annealing and cooling down to room temperature, RHEED patterns of the resultant sample are shown in Fig. 1(d). The distinct spots from specular and off specular reflections with the well developed streak patterns confirm excellent CCO thin film crystallinity and flat terraces. In addition, smooth surface morphology is corroborated by the AFM imaging shown in Fig. 1(e); the obtained average surface roughness is below 60 pm for a 1 $\mu$m by 1 $\mu$m scan.   

It is of critical importance to ensuring the (1 1 1) growth direction is maintained throughout the sample and determining the film thickness, thus we performed the X-ray diffraction 2$\theta$-$\omega$ scans on CCO thin films as well as on the pure AlO substrate, together with a X-ray reflectivity scans on the film using Cu K$_{\alpha}$ radiation. As shown in Fig. 2, the expected CCO (2 2 2), (3 3 3) and (4 4 4) Bragg reflections are clearly seen, which testify that the sample is (1 1 1) orientated. The lattice constant calculated from the peak position is 8.33 \AA, which is in excellent agreement with the bulk value\cite{Menyuk}. Based on the XRD pattern, no impurity phases are detected. In addition, the inset in Fig 2 displays the X-ray reflectivity, yielding a film thickness of about 25 nm according to the period of the thickness fringes that provides additional evidence for the sample flatness.       

Due to the well-known persistent multivalency problem in the thin film growth of spinel oxides, the proper valency of the Cr and Co ions must be confirmed. Fig. 3(a) and 3(b) show XPS measurements taken on Cr and Co 2p core levels. As seen, Cr has two peaks at about 577 eV and 587 eV which correspond to the 2p$_{3/2}$ and 2p$_{1/2}$ peaks indicating Cr is in the trivalent state.\cite{Wagner, Lei} The 2p$_{3/2}$ and 2p$_{1/2}$ peaks of Co are at 781 eV and 797 eV indicating Co is in the divalent state\cite{Wagner, Lei}. In addition, no extra peaks from Cr$^{2+}$,  Cr$^{4+}$ or Co$^{3+}$ are observed, which excludes the possibility of a multivalent charge state. 

To further elucidate that the CCO thin film maintains the normal spinel structure without cation disorder, XRMS  experiments were carried out at beamline 4-ID-C of the Advanced Photon Source (APS) in Argonne National Laboratory. Left (I$^+$) and right (I$^-$) polarized soft x-rays with an incident angle of 15$^{\circ}$ were tuned to the $\it{L}$ edges of Cr and Co and recorded in scattering mode\cite{Freeland}. In the XRMS mode, the difference between I$^+$ and I$^-$  near the absorption edge represents the contribution to the scattering amplitude from uncompensated magnetic moments of a specific chemical  element\cite{Kao}, while the sum I$^+$ + I$^-$ is connected to the charge state\cite{Lee}. Figures 3(c) and 3(d) show the total reflectivity intensity (I$^+$ $+$ I$^-$) of both Cr and Co as a function of incident photon energy. As seen, the lineshapes of each element measured at 15 K and 80 K are almost unchanged. The positions of the main absorption edges $\it{L}$$_3$ and $\it{L}$$_2$ together with the satellite peaks are almost identical to the previous reported study\cite{Chopdekar}, which implies all of the Cr$^{3+}$ are in the octahedral sites while all the Co$^{2+}$ sit in the tetrahedral sites.    

Next we discuss the magnetic properties of the CCO thin film. To this end, temperature dependent magnetization was measured while cooling in an applied magnetic field of 0.2 T by SQUID. As seen in Fig. 4(a), the ferrimagnetic phase transition occurs at a T$_c$ around 95 K. As the temperature further decreases, the collinear to incommensurate spiral ferrimagnetic phase transition\cite{Tomiyasu} takes place at T$_s$ $\approx$ 22 K. Element-specific spin alignments were further investigated by XRMS spectra on Cr and Co $\it{L}$$_{2,3}$ absorption edges at 15 K and 80 K in different applied fields from 5 T to 0.1 T. All of the XRMS data have been normalized by using the corresponding total reflectivity intensity. As presented in Fig. 4(b) and 4(c), both Cr and Co exhibit significant XRMS signal in the vicinity of their absorption edges. The maximal XRMS signal at Cr $\it{L}$$_3$ edge is about +20\% while that of Co is around -80\%, which indicate strong ferromagnetic ordering of the moments on each type of metal ions, respectively. Note, the sign of the maximal XRMS signal is opposite for Cr and Co due to the ferrimagnetic nature of this material, which implies the overall spin orientation of Cr is antiparallel to those of the Co ions. In Fig. 4(d), the magnitudes of the maximal XRMS values of Cr and Co  at 15 K and 80 K are plotted as a function of applied magnetic field showing a saturation field of approximately 3 T. This is much larger than the reported bulk values\cite{Tomiyasu, Yamasaki, Lei} which are typically less than 0.6 T. This difference is attributed to the fact that in the bulk the magnetic field is commonly applied along the CCO [0 0 1] direction which is the spin easy axis.  


In summary, we have successfully fabricated high quality (1 1 1)-oriented normal spinel CCO thin films on AlO (0 0 0 1) substrate. The structural, electronic and magnetic properties are investigated by a combination of RHEED, XRD, XPS, XRMS and SQUID measurements. No disorder of the cation distribution or multivalency is observed. Magnetic measurements confirm the ferrimagnetic behavior. Since all the trivalent Cr are situated in the octahedral sites, they compose magnetic Kagom$\acute{e}$ planes naturally layered  along this orientation. The presented results pave a way for fabricating 2D-isolated Kagom$\acute{e}$ heterostructures in which a plethora of novel quantum phenomena are expected.


The authors acknowledge M. Hawkridge for the assistances on the XRR measurement. JC deeply acknowledges numerous fruitful discussions with  D. Khomskii and G. Fiete. J.C. was supported by the DOD-ARO under Grant No. 0402-17291. Work at the Advanced Photon Source, Argonne is supported by the U.S. DOE under Grant No. DEAC02Â06CH11357.

\newpage

\begin{figure}[t]\vspace{-0pt}
\includegraphics[width=1\textwidth]{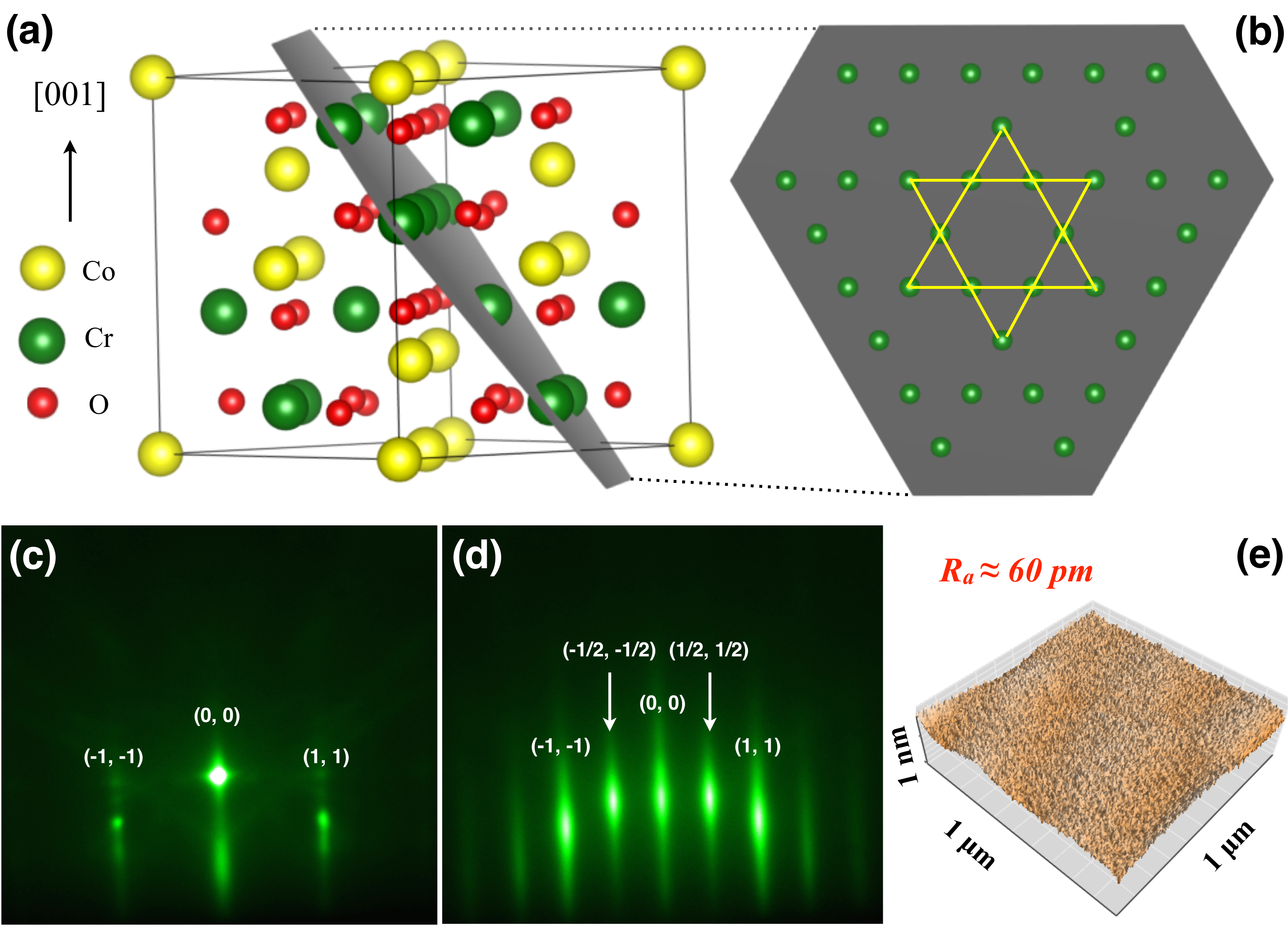}
\caption{\label{} (a) Schematic crystal structures of the CCO unit cell. Cr$^{3+}$ ions along the (1 1 1) orientation compose a Kagom$\acute{e}$ lattice plane. (b) Detailed 2D Kagom$\acute{e}$ lattice plane. The typical corner-shared triangular structure is outlined by yellow lines. (c) RHEED patterns of  AlO substrate. (d) RHEED patterns of CCO thin films after cooling down to room temperature. Note, the incident electron beam of the RHEED is fixed along the [1 $\bar{1}$ 0 0] direction of the substrate. (e) AFM image of the film surface after growth.}
\end{figure}

\newpage

\begin{figure}[t]\vspace{-0pt}
\includegraphics[width=1\textwidth]{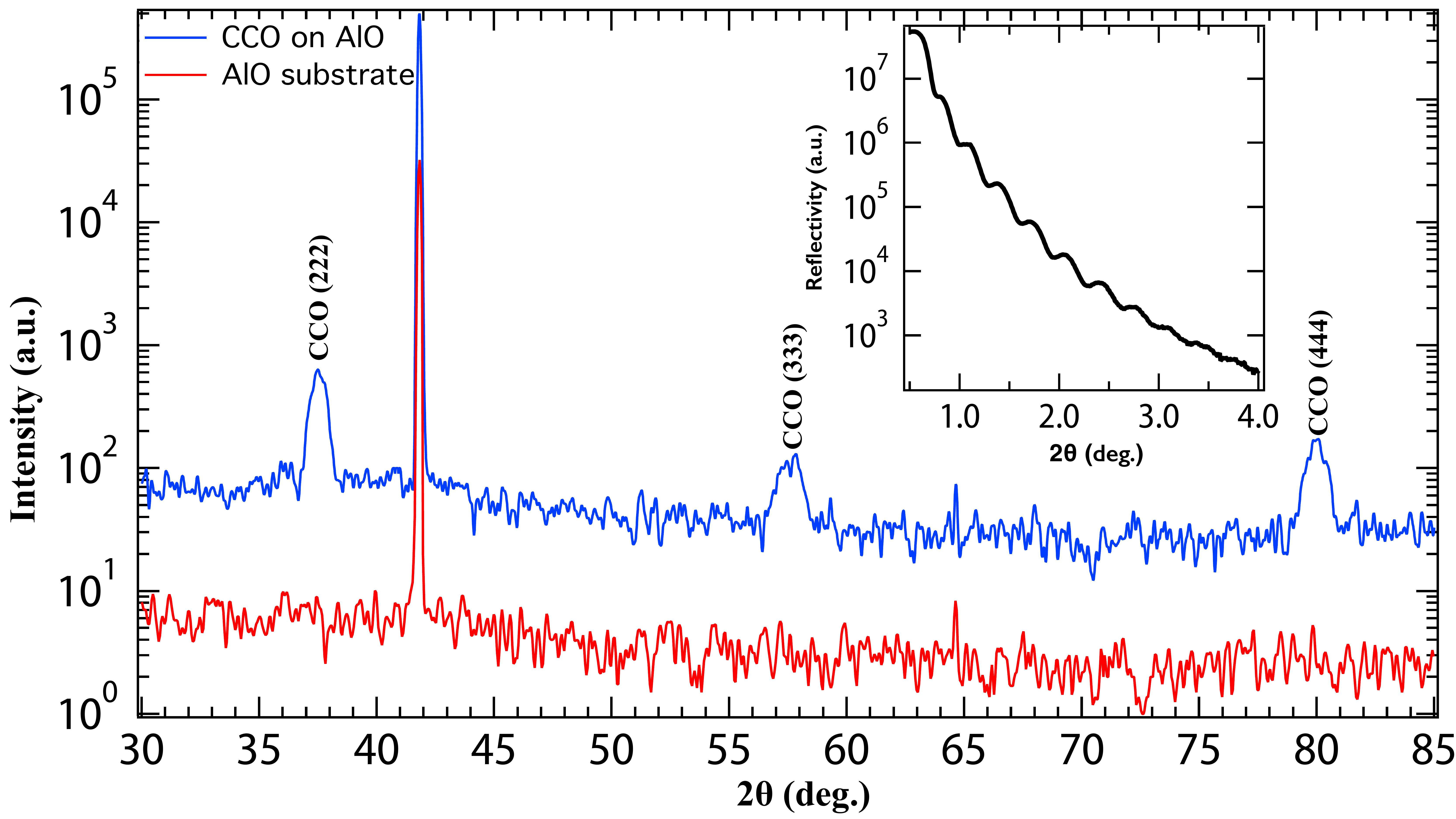}
\caption{\label{} X-ray diffraction of CCO thin films and the AlO substrate. Film peaks are labeled on the graph. The sharp peaks belong to the AlO (0 0 0 1) substrate. Note, the lattice constant obtained is 8.34 \AA, which equals the bulk value. The inset is the X-ray reflectivity data of the same sample. Film thickness calculated according to the Kiessig fringes is about 25 nm.}
\end{figure}

\newpage

\begin{figure}[t]\vspace{-0pt}
\includegraphics[width=1\textwidth]{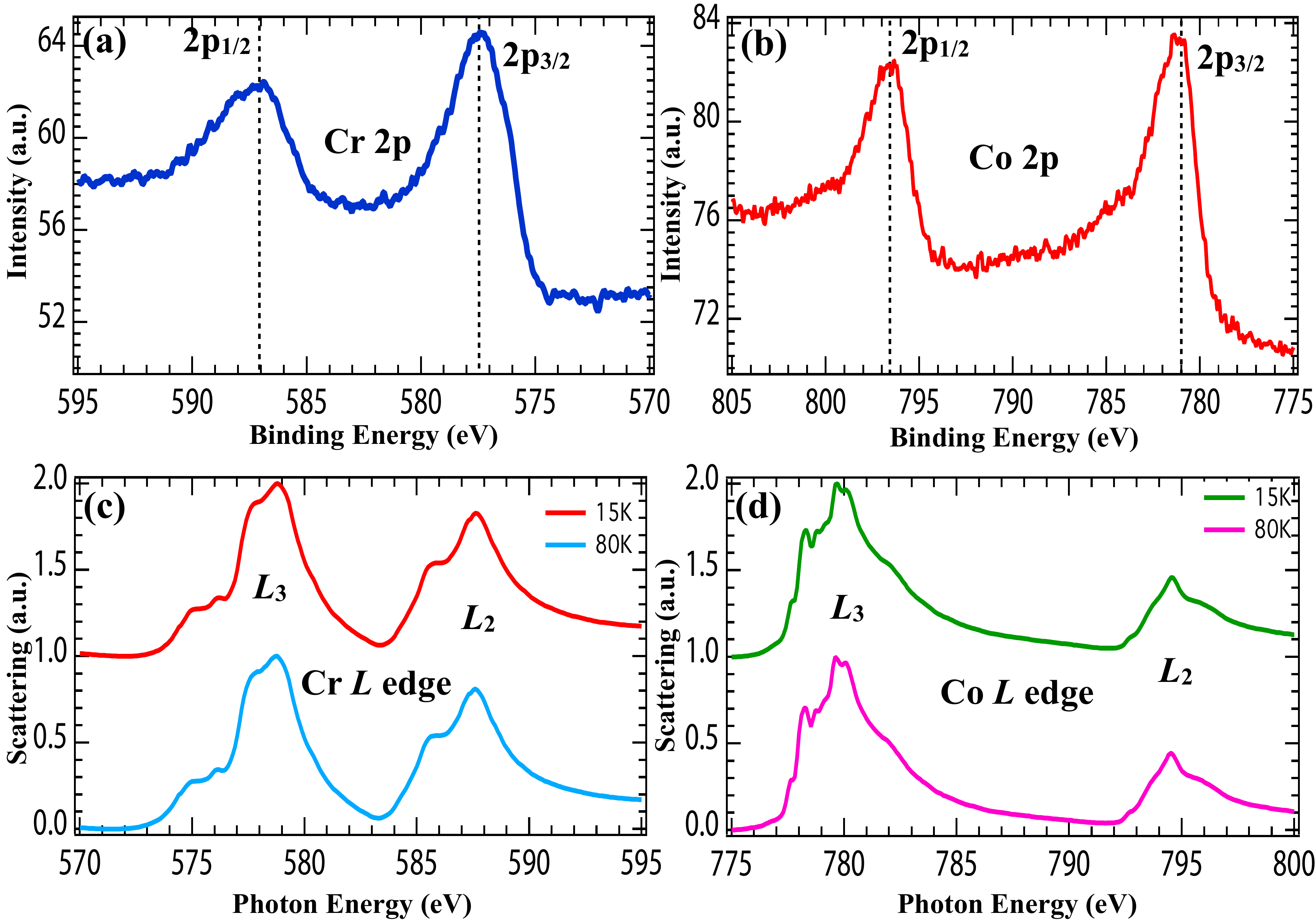}
\caption{\label{} (a)-(b) Core level XPS data (Mg anode) of (a) Cr 2p and (b) Co 2p. (c)-(d) X-ray scattering spectra measured at 15 K and 80 K on the $\it{L}$ edge of (c) Cr and (d) Co.}
\end{figure}

\newpage

\begin{figure}[t]\vspace{-0pt}
\includegraphics[width=1\textwidth]{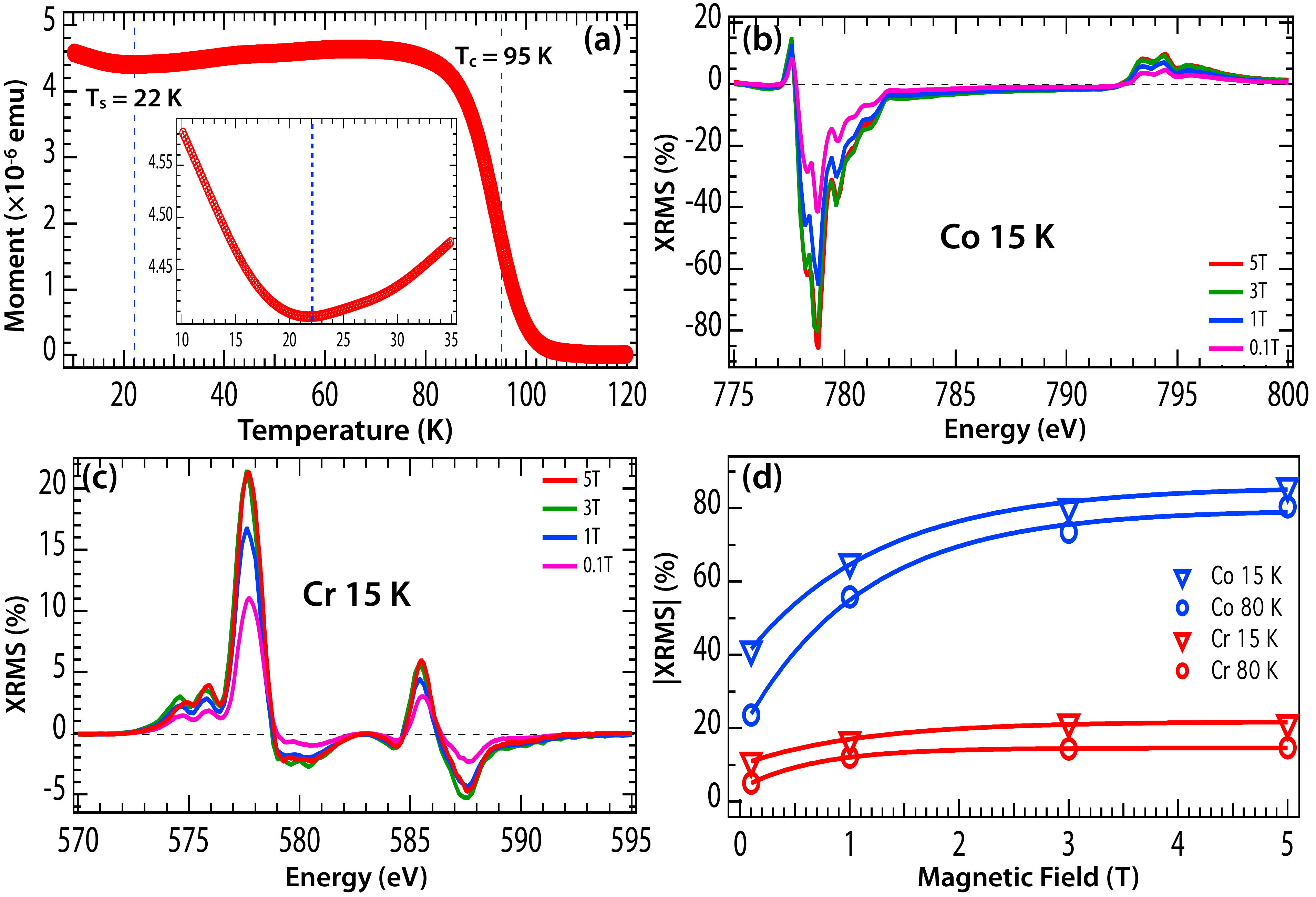}
\caption{\label{} (a) Temperature-dependent magnetization curves of CCO in an applied field of 0.2 T  along the [1 $\bar{1}$ 0 0] direction of the substrate. The inset is a magnified plot in the vicinity of the second transition point. (b)-(c) XRMS data on the L edge of (b) Co and (c) Cr measured with different applied magnetic fields at 15  K. The color series of red, green, blue and pink stands for field strength of 5 T, 3 T, 1 T and 0.1 T, respectively. (d) Absolute values of Co and Cr XRMS main peaks as a function of applied field and temperature.} 
\end{figure}


\begin{thebibliography}{999}

\bibitem{Guo}
H. M. Guo, and M. Franz, Phys. Rev. B \textbf{80}, 113102(2009)

\bibitem{Hu}
X. Hu, A. R\"{u}egg, and G. A. Fiete, Phys. Rev. B \textbf{86}, 235141(2012)

\bibitem{Han}
T. H. Han, J. S. Helton, S. Chu, D. G. Nocera, J. A. Rodriguez-Rivera, C. Broholm, and Y. Lee, Nature \textbf{492}, 406-410(2012)  

\bibitem{Punk}
M. Punk, D. Chowdhury, and S. Sachdev, Nat. Phys. \textbf{10}, 289-293(2014)

\bibitem{Pollmann}
F. Pollmann, P. Fulde, and K. Shtengel, Phys. Rev. Lett. \textbf{100}, 136404(2008)

\bibitem{Liu}
G. Liu, P. Zhang, Z. Wang, and S. Li, Phys. Rev. Lett. \textbf{79}, 035323(2009)

\bibitem{Petrescu}
A. Petrescu, A. A. Houck, and K. Le Hur, Phys. Rev. A \textbf{86}, 053804(2012)

\bibitem{Yu}
S. Yu and J. Li, Phys. Rev. B \textbf{85}, 144402(2012)

\bibitem{Yunling}
Y. Liu, V. Ch. Kravtsov, D. A. Beauchamp, J. F. Eubank, and M. Eddaoudi, J. Am. Chem. Soc. \textbf{127}, 7266-7267(2005)

\bibitem{Emily}
E. A. Nytko, J. S. Helton, P. M\"{u}ller, and D. G. Nocera, J. Am. Chem. Soc. \textbf{130}, 2922-2923(2008)

\bibitem{Gibert}
M. Gibert, P. Zubko, R. Scherwitzl, J. Iniguez, and J. M. Triscone, Nat. Mater. \textbf{11}, 195(2012) 

\bibitem{Sri}
S. Middey, D. Meyers, M. Kareev, E. J. Moon, B. A. Gray, X. Liu, J. W. Freeland, and J. Chakhalian, Appl. Phys. Lett. \textbf{101}, 261602(2012)

\bibitem{Ben}
B. Gray, H. Lee, J. Liu, J. Chakhalian, and J. W. Freeland, Appl. Phys. Lett. \textbf{97}, 013105(2010)

\bibitem{Yahiro}
H. Yahiro, H. Tanaka, Y. Yamamoto, and T. Kawai, Solid State Commun. \textbf{123}, 535-538(2002)

\bibitem{Menyuk}
N. Menyuk, K. Dwight, and A. Wold, J. Phys. France \textbf{25}, 528-536(1964)

\bibitem{Choi}
Y. J. Choi, J. Okamoto, D. J. Huang, K. S. Chao, H. J. Lin, C. T. Chen, M. van Veenendaal, T. A. Kaplan, and S-W. Cheong, Phys. Rev. Lett. \textbf{102}, 067601(2009)

\bibitem{Tomiyasu}
K. Tomiyasu, J. Fukunaga, and H. Suzuki, Phys. Rev. B \textbf{70}, 214434(2004)

\bibitem{Yamasaki}
Y. Yamasaki, S. Miyasaka, Y. Kaneko, J. P. He, T. Arima, and Y. Tokura, Phys. Rev. Lett. \textbf{96}, 207204(2006)

\bibitem{Singh}
K. Singh, A. Maignan, C. Simon, and C. Martin, Appl. Phys. Lett. \textbf{99}, 172903(2011)

\bibitem{Tsurkan}
V. Tsurkan, S. Zherlitsyn, S. Yasin, V. Felea, Y. Skourski, J. Deisenhofer, H. A. Krug von Nidda, J. Wosnitza, and A. Loidl, Phys. Rev. Lett. \textbf{110}, 115502(2013)

\bibitem{Kaplan}
T. A. Kaplan, and N. Menyuk, Philos. Mag. \textbf{87}, 3711-3785(2007)  

\bibitem{Aria}
A. Yang, Z. Chen, X. Zuo, J. Kirkland, C. Vittoria, and V. G. Harris, Appl. Phys. Lett. \textbf{86}, 252510(2005)

\bibitem{Ma}
J. X. Ma, D. Mazumdar, G. Kim, H. Sato, N. Z. Bao, and A. Gupta, J. Appl. Phys. \textbf{108}, 063917(2010)

\bibitem{Ulrike}
U. L\"{u}ders, M. Bibes, J. Bobo, M. Cantoni, R. Bertacco, and J. Fontcuberta, Phys. Rev. B \textbf{71}, 13441992005)

\bibitem{Matzen}
S. Matzen, J.-B. Moussy, R. Mattana, K. Bouzehouane, C. Deranlot, F. Petroff, J. C. Cezar, M.-A. Arrio, Ph. Sainctavit, C. Gatel, B. Warot-Fonrose, and Y. Zheng, Phys. Rev. B \textbf{83}, 184402(2011)

\bibitem{Kareev}
M. Kareev, S. Prosandeev, B. Gray, J. Liu, P. Ryan, A. Kareev, E. J. Moon, and J. Chakhalian, J. Appl. Phys. \textbf{109}, 114303(2011)

\bibitem{Moussy}
J. Moussy, J. Phys. D: Appl. Phys. \textbf{46}, 143001(2013)

\bibitem{Wagner}
C. D. Wagner, W. M. Riggs, L. E. Davis, J. F. Moulder, and G. E. Muilenberg, \textit{Handbook of X-ray Photoelectron Spectroscopy} (Perkin-Elmer, Eden Prairie, 1979)

\bibitem{Lei}
S. Lei, L. Liu, C. Wang, X. Shen, C. Wang, D. Guo, S. Zeng, B. Cheng, Y. Xiao, and L. Zhou, CrystEngComm \textbf{16}, 277(2014)

\bibitem{Freeland}
J. W. Freeland, J. J. Kavich, K. E. Gray, L. Ozyuzer, H. Zheng, J. F. Mitchell, M. P. Warusawithana, P. Ryan, X. Zhai, R. H. Kodama, and J. N. Eckstein, J. Phys: Condens. Matter \textbf{19}, 315210(2007)

\bibitem{Kao}
C. Kao, J. B. Hastings, E. D. Johnson, D. P. Siddons, G. C. Smith, and G. A. Prinz, Phys. Rev. Lett. \textbf{65}, 373(1990)

\bibitem{Lee}
D. R. Lee, and S. K. Sinha, Phys. Rev. B \textbf{68}, 224409(2003)

\bibitem{Chopdekar}
R. V. Chopdekar, M. Liberati, Y. Takamura, L. F. Kourkoutis, J. S. Bettinger, B. B. Nelson-cheeseman, E. Arenholz, A. Doran, A. Scholl, D. A. Muller, and Y. Suzuki, J. Magn. Magn. Mater. \textbf{322}, 2915-2921(2010)


\end{thebibliography}
\end{document}